\documentclass[aps,prl,twocolumn,superscriptaddress]{revtex4}
\usepackage{epsfig}
\begin{document}

\title[Short title for running header]{Proximity effect in ultrathin Pb/Ag
multilayers within the Cooper limit}
\author{O. Bourgeois$^{1,3}$, A.Frydman$^{2}$ and R.C. Dynes$^{}$}
\affiliation{$^{}$The Department of Physics, University of
California, San Diego, La
Jolla, CA 92093\\
$^{2}$Department of Physics, Bar Ilan university, Ramat Gan,
52900, Israel\\
 $^{3}$ CRTBT/CNRS 25 Avenue des Martyrs BP 166 38042 Grenoble Cedex 9
France}


\begin{abstract}
We report on transport and tunneling measurements performed on
ultra-thin Pb/Ag (strong coupled superconductor/normal metal)
multilayers evaporated by quench condensation. The critical
temperature and energy gap of the heterostructures oscillate with
addition of each layer, demonstrating the validity of the Cooper
limit model in the case of multilayers. We observe excellent
agreement with a simple theory for samples with layer thickness
larger than 30\AA . Samples with single layers thinner than 30\AA\
deviate from the Cooper limit theory. We suggest that this is due
to the ''inverse proximity effect'' where the normal metal
electrons improve screening in the superconducting ultrathin layer
and thus enhance the critical temperature.
\end{abstract}

\maketitle

The superconducting proximity effect is a well known phenomenon
and has drawn a lot of interest both from the fundamental and the
practical points of view. In a high transmission contact between a
superconductor and a normal metal the superconducting
wave-function varies smoothly across the interface causing a
suppression of the pair amplitude in the superconductor and an
enhancement of pairing on the normal side. The characteristic
distance in which superconductivity ``leaks'' into the normal
region is the
normal state coherence length $\xi _{N}=\sqrt{\text{%
h{\hskip-.2em}\llap{\protect\rule[1.1ex]{.325em}{.1ex}}{\hskip.2em}%
}D/2\Pi kT}$, where D is the diffusivity, and the length-scale at
which superconductivity is suppressed in the superconducting side
is the superconducting coherence length, $\xi _{S}$.

In the Cooper limit \cite{cooper} (in which both the superconductor and the
normal metal layers are smaller than the characteristic coherence lengths $%
\xi _{S}$ and $\xi _{N}$ respectively) the behavior of an S-N
bilayer is well described within the framework of the de Gennes
model \cite{de Gennes,deutscher}. The electrons experience an
average paring interaction (average between the two materials) and
$T_{C}$ can be described within the BCS weak coupling form adapted
to the Cooper limit proximity effect in the following way :

\begin{equation}
kT_{C}=1.13\text{%
h{\hskip-.2em}\llap{\protect\rule[1.1ex]{.325em}{.1ex}}{\hskip.2em}%
}\omega _{D}e^{-\frac{1}{N(0)V}}  \label{BCS}
\end{equation}

\[
\lbrack N(0)V]_{S+N}=\frac{d_{S}[N(0)V]_{S}+dN[N(0)V]_{N}}{d_{S}+d_{N}}
\]%
here $\omega _{D}$ is the debye frequency, N(0)V is the pairing
interaction (N(0) being the density of states at the Fermi level
and V the pairing interaction), and d$_{S,N}$ is the thickness of
the superconducting and normal metal films respectively. This
model assumes that the debye frequencies in the two metals are
similar and it is based on the
understanding that the electrons spend $\frac{N_{N}d_{N}}{%
(N_{N}d_{N}+N_{S}d_{S})}$ of their time in the normal metal and $\frac{%
N_{S}d_{S}}{(N_{N}d_{N}+N_{S}d_{S})}$ in the superconductor
($N_{N}$ and $N_{S}$ are the density of states in the normal metal
and in the superconductor respectively).

The above picture is modified in the case of strong coupled superconductors
where $T_{C}$ is given by the McMillan expression \cite{mcmillen} :

\begin{equation}
T_{C}=\frac{\omega _{D}}{1.45}\exp -[\frac{1.04(1+\lambda )}{\lambda -\mu
^{\ast }(1+0.62\lambda )}]  \label{strong coupling}
\end{equation}

here $\lambda $ is the dimensionless electron-phonon coupling
coefficient and $\mu ^{\ast }$ is the effective coulomb repulsion.
In a normal-metal/strong coupled superconductor within the Cooper
limit one can expect $\lambda $ in equation \ref{strong coupling}
to be replaced by an average electron-phonon coupling
\cite{silbert}:

\begin{equation}
\lambda _{S+N}=\frac{\lambda _{S}d_{S}+\lambda _{N}d_{N}}{d_{S}+d_{N}}
\label{lamda}
\end{equation}

The proximity effect in the Cooper limit has been observed
experimentally in numerous systems of superconductor/normal metal
bilayers. In this paper we study systems of multilayers of
superconductor/normal metal (Pb/Ag) in which the total thicknesses
of the metals are comparable to the relevant coherence lengths.
The transport and tunneling experimental results confirm that the
basic idea of the Cooper limit notion, i.e. that the electrons
experience an average pairing, is valid also in a multilayer
sample. We observe good \emph{quantitative} agreement with theory
as long as the thickness of the layers is larger than $\sim
30$\AA\ while thinner layers deviate from the theoretical
predictions. We suggest that the simple picture is modified in
ultra-thin layers due to effects of screening of e-e interactions
(the inverse proximity effect as previously reported
\cite{inverse}).

The Pb and Ag layers were prepared by quench condensation, i.e. evaporation
on a cryogenically cooled substrate within the measurement apparatus \cite%
{strongin}. This allows \textit{in-situ} sequential depositions
under UHV conditions and simultaneous transport and tunneling
measurements on a single sample. This method has several essential
advantages for the study of the superconducting proximity effect
\cite{prox}. Since Pb and Ag are immiscible, the alloying of these
two materials is very improbable even at room temperatures.
Alloying becomes practicably negligible when the samples are
quench condensed and are kept below T=10K throughout the entire
experiment (sample growth and measurement). In addition, the
extremely clean environment of the experiment leads to very high
quality interfaces between the superconducting layers and the
normal metal layers. Such clean interfaces provide ideal
conditions for proximity effects.

The geometry of the samples used in this study is shown in the
insert of figure 1. We begin by evaporating two strips of Al onto
a room temperature polished glass substrate. These strips are used
both as voltage leads and as the base electrode for tunneling
measurements. Then we allow a native oxide barrier to grow on the
Al for 20 minutes. We connect metallic leads to the substrate and
place it in an evaporation chamber which is then pumped out and
immersed in liquid helium. A 10\AA\ Ge adhesion layer is then
quench condensed across the Al strips, permitting the evaporation
of a continuous layer of Pb as thin as 5\AA\ \cite{strongin}. The
thicknesses are monitored by a calibrated quartz crystal situated
in the chamber. Multilayers, made of base units of Pb/Ag bilayers
having different thicknesses, are quench condensed across the Al
strips. This configuration allows the simultaneous measurements of
transport properties of the multilayers (thus determining the
critical temperature), and tunneling into the multilayers (for
determination of the energy gap). We performed these measurements
at incremental stages of the evaporation of each of the Pb or Ag
layers. All measurement were performed in a screened room using
standard AC methods.

\begin{figure}\centering
\epsfxsize8.5cm\epsfbox{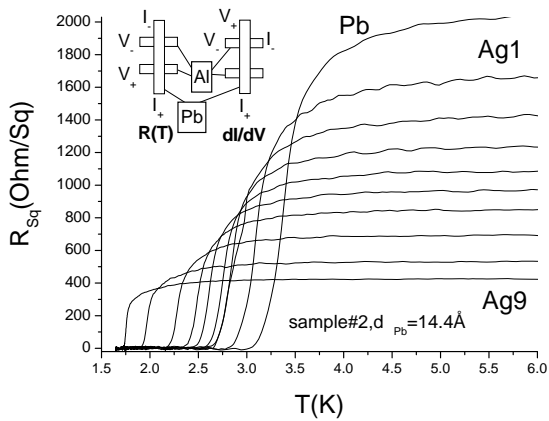} \vskip -4.5truecm\caption
{Resistance per square versus temperature of a Pb/Ag bilayer
(sample \#2). The thickness of Ag varies from 0 \AA\ (Pb) to 14.4
\AA\ (Ag9). Each evaporation corresponds to approximately 1.6 \AA
. The $T_{C}$ decreases monotonically as normal state resistance
is decreased. Insert: Illustration of the sample geometry.
 }\end{figure}

In this paper we present data on three Pb/Ag
systems having different base unit thickness: Sample \#1 where
$d_{pb}=d_{Ag}\approx $32\AA , sample \#2
where $d_{pb}=d_{Ag}\approx $15\AA , and sample \#3 where $%
d_{pb}=d_{Ag}\approx $9\AA . Each of these unit bilayers is within
the cooper limit; $\xi _{N}\;$of the Ag at low temperatures is of
the order of 1 $\mu m$ and $\xi _{S}$ of the amorphous Pb (which
is in the dirty limit) at low temperature is given by $\xi
^{\prime }=\sqrt{\xi _{0}l}$ where $\xi _{0} $ is the clean limit
coherence length (for Pb this is 800\AA ) and \emph{l} is the mean
free path which is approximately an interatomic distance, about
2.7\AA\ in our samples \cite{remark2}. These values yield a zero
temperature coherence length, $\xi ^{\prime }\approx 46$\AA ,\
which is larger than each one of our Pb layers. Furthermore, as
the temperature increases the coherence length increases until it
diverges at $T=T_{C}.$
\begin{figure}\centering
\epsfxsize8.5cm\epsfbox{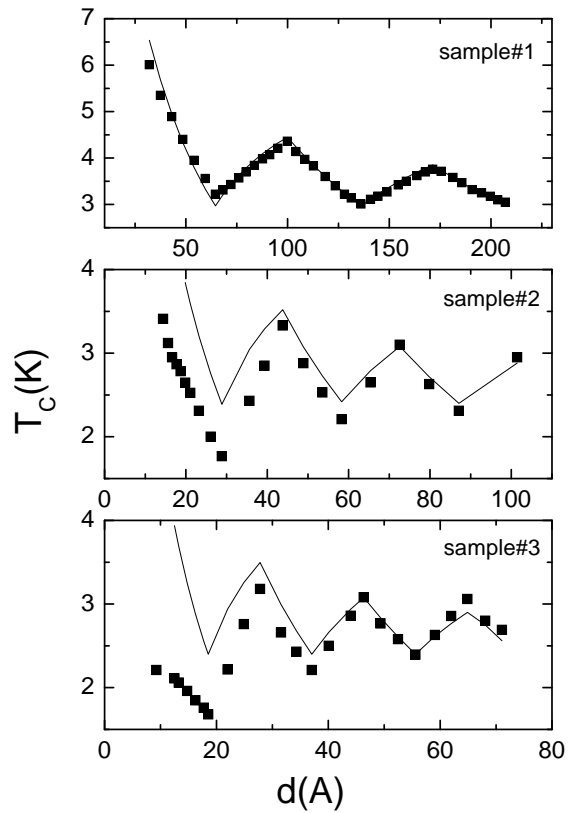} \caption {Superconducting
critical temperatures extracted from the R(T) measurements in the
case of three different thickness of the Pb/Ag layer. The
thicknesses of the Pb and Ag layers in the bilayer are 32.5 \AA\
(top graph), 14.4 \AA\ (middle graph) and 9.3 \AA\ (bottom graph).
The solid lines are the theoretical fits using equations 2 and 3.
}\end{figure}

We begin by presenting the results for the initial bilayer. Figure 1 shows
the resistance versus temperature of sample \#2 as a layer of 14.5 \AA\ %
thick Ag is grown in steps of $\approx $1.55\AA\ on top of a
14.4\AA\ thick initial Pb layer. Such measurements are used to
extract $T_{C}$ by taking the midpoint of the transition
(R$_{N}$/2) \cite{remark1}. It is seen that the initial Pb layer
has a $T_{C}$ which is smaller than the bulk value of 7.2K. This
is typical for very thin superconducting layers which are known to
exhibit a superconductor - insulator transition as a function of
film
thickness (and R$_{sq}$) \cite{graybeal,white,goldman}. It has been argued %
\cite{uniform,lee} that as R$_{sq}$ (the disorder) increases, the electronic
screening is reduced. Coulomb interactions are then enhanced leading to a
weakening of the superconducting coupling. This explains the decrease of $%
T_{C}$ in the Pb films when the thickness is reduced from 30\AA\ to 9\AA\ as
is seen in figure 2. The enhancement of Coulomb interactions for high R$_{sq}
$ is also evident in figure 3 which depicts the dI/dV-V measurements
(proportional to the density of states) of a 9\AA\ thick Pb film with
increasing layers of Ag, taken at temperatures above $T_{C}$. It is seen
that in the ultrathin Pb layer the density of states at the Fermi level is
relatively low. The tunneling conductance exhibits a strong voltage (energy)
dependence demonstrating that this is a strongly coulomb correlated system.
The density of states dramatically increases when layers of Ag are
evaporated on top of the Pb, presumably due to enhancement of screening and,
hence, suppression of the Coulomb interactions \cite{uniform}.

As subsequent layers of Ag are added (and R$_{sq}$ is reduced as
$1/d_{Ag}$ ) $T_{C}$ is decreased even further due to the
proximity effect as evident in figure 1. Since the bilayer is in
the cooper limit, $T_{C}$ is indeed expected to decrease as a
function of the normal metal thickness. We note, however, that
this is not always the case. We have shown \cite{inverse} that
when the initial Pb layer is extremely thin (thinner than that of
sample 3) the first added Ag sub-layers cause $T_{C}$ to\textit{\
increase} in what seems to be an inverse proximity effect . This
behavior was interpreted as being due to the fact that the Ag
layers enhance screening in the system thus suppressing the e-e
interactions which are responsible for the $T_{C}$ suppression.
This trend opposes the usual proximity effect and becomes more
important in thinner superconducting films.

\begin{figure}\centering
\epsfxsize8.5cm\epsfbox{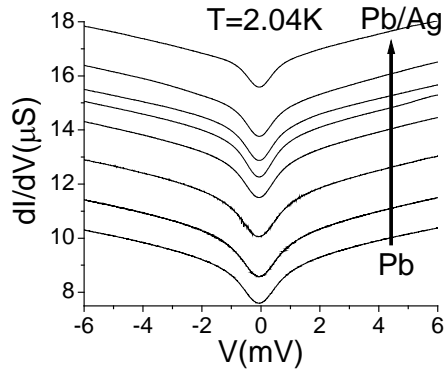} \caption {dI/dV versus V
(proportional to the density of states as a function of energy) at
temperatures above $T_{C}$ (2.04K) of an ultrathin Pb layer (10
\AA ) with an increasing Ag (up to 10 \AA ) layer on top.
}\end{figure}

The evolution of $T_{C}$ as more Pb and Ag layers are added is shown in
figure 2 for the three samples. It is seen that $T_{C}$ oscillates when we
change between Pb and Ag depositions reaching a local minimum or maximum at
the completion of each layer. The oscillation amplitude decreases with
increasing number of bilayers and T$_{C}$ appears to approach a value of $%
\approx 3K$ in all three samples. The oscillation behavior is also
observed in the energy gap measurements which are illustrated in
figure 4 for sample \#1. The dI/dV curves were measured at T=1.65K
well below the $T_{C}$ of the multilayer. Note that despite that
fact that tunneling is performed into the bottom Pb layer, the
results are still affected by adding the 6th layer on top. This
clearly demonstrates that within the cooper limit the tunneling
electrons probe the \emph{entire} sample and the superconducting
parameters are determined by the mean properties of \textit{all}
the layers in the multilayer. Figure 4 also shows the ratio
$\frac{2e\Delta }{kT_{C}}$ as a function of the number of layers.
The superconducting gap was evaluated from a classical temperature
dependent BCS fit. It is seen that this ratio crosses over
from a value of 4.7 (which is close to the strong coupling limit of $\sim $%
4.8) to 3.6 (which is close to the week coupling limit of $\sim
$3.5). The fact that this ratio is relatively constant as we add
the last few layers gives us further confidence that all our
multilayers are within the Cooper limit.

We have attempted to fit the curves in figure 2 to the proximity
effect modified McMillan expression of equation \ref{strong
coupling} using an average $\lambda $ for the Cooper limit
(equation \ref{lamda}) in the spirit of the de-Gennes model. These
are shown in figure 2 were we used the known values for Pb:
$\lambda _{S}=1.55,\omega _{D}=105K,\mu ^{\ast }=0.11$ and taking
a low electron-phonon coupling for Ag, $\lambda _{N}=0.2,$ without
any fitting parameters. It is seen that the experimental results
are in
excellent agreement with this model for sample \#1 (where $%
d_{pb}=d_{Ag}\approx $32\AA ). We note that the values for
$\lambda ,\omega _{D}$ and $\mu ^{\ast }$ are not adjusted, but
accepted values and so the agreement is quite impressive. However,
when we used the same parameters for the other samples with
thinner base layers, we get substantial deviations from the model
for the first few layers (figure 2). In fact,  in sample \#1 (and
in the theory) the oscillations are superimposed on a global
decrease of $T_{C},$ in sample \#2 the background is roughly
constant and in sample \#3 T$_{C}$ globally \textit{increases}
with the number of bilayers. We argue that this behavior is an
extension of the inverse proximity effect observed in the
ultrathin bilayers \cite{inverse}. The large Coulomb interactions
characteristics of these thin samples are continuously weakened
with increasing numbers of Pb/Ag bilayers, thus, as more layers
are added, we approach the Cooper limit proximity effect theory.

\begin{figure}\centering
\epsfxsize8.5cm\epsfbox{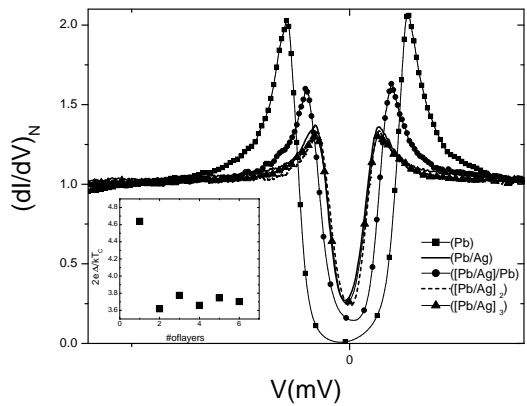} \vskip -4.5truecm\caption
{dI/dV-V, normalized to the normal state conductance, for sample
\#1 taken at T=1.65K for each Pb and Ag sub-layer. The
oscillations
of the gap with thickness are easily observed. The inset shows $\frac{%
2e\Delta }{kT_{C}}$ as a function of the number of layers. Note
that this ratio is nearly constant for the last few layers.
}\end{figure}

In summary, we have shown that the de Gennes considerations
regarding the Cooper limit are valid for the case of multilayers
until the layers become very thin. Using a simple model we were
able to describe very accurately the variation of the critical
temperature and the energy gap as a function of the number of
layers. We showed that for very thin layers the model breaks down
indicating that other physics is involved. We suggest that this
breakdown is due to the effect of strong Coulomb interactions in
the superconducting ultrathin films. These interactions are
screened out as normal metal is added to the superconductor
resulting in an inverse proximity effect which must also be taken
into account in the thinnest layers.

We gratefully thank O. Naaman and W. Teizer for fruitful
discussions. The work was supported by the Israel-US Binational
Foundation \#1999332 and by the NSF grant \#DMR0097242.


\begin{thebibliography}{99}
\bibitem{cooper} L. Cooper, Phys. Rev. Lett. \textbf{6}, 689 (1961).

\bibitem{de Gennes} P.G. de Gennes, Rev. Mod. Phys, \textbf{36}, 225 (1964).

\bibitem{deutscher} G. Deutscher and P.G. de Gennes in superconductivity,
edited by Parks, Marcell Dekker Inc. (New York 1969).

\bibitem{mcmillen} W.L. McMillan, Phys. Rev. \textbf{167}, 331(1968).

\bibitem{silbert} W. Silbert Phys. Rev. \textbf{B12} 4870 (1975).

\bibitem{inverse} O. Bourgeois, A. Frydman and R.C. Dynes, Phys. Rev. Lett.,
\textbf{88}, 186403 (2002).


\bibitem{strongin} M. Strongin, R. Thompson, O. Kammerer and J. Crow, Phys.
Rev. \textbf{B1}, 1078 (1970).

\bibitem{prox} A. Frydman and R.C. Dynes, Solid State Commun. \textbf{110},
485 (1999).

\bibitem{remark2} This value was calculated from the density of Pb and is
also consistent with the conductivity measures for 1-2 monoatomic thickness
of our films.

\bibitem{remark1} The relatively broad transition is a consequence of
superconductor fluctuations which introduce some uncertainty when
evaluating T$_{C}$. This can be overcome by fitting the curves
using the Aslamazov Larkin fluctuation theory (L.G. Aslamasov and
A.I. Larkin, Phys. Lett. \textbf{26A}, 238 (1968)). Such fits
yield T$_{C}$s which are very close to the midpoint of the
transition as shown in ref 6.

\bibitem{graybeal} J.M. Graybeal and M.R. Beasley, Phys. Rev. \textbf{B29},
4167 (1984).

\bibitem{white} A.E. White, R.C. Dynes and J.P. Garno, Phys. Rev. \textbf{B33%
}, 3549 (1986).

\bibitem{goldman} A.M. Goldman and N. Markovic, Physics Today \textbf{51},
39 (1998) and references within.

\bibitem{uniform} J.M. Valles Jr., R.C. Dynes and J.P. Garno, Phys. Rev.
\textbf{B40}, 6680 (1989).

\bibitem{lee} P.A. Lee and T.V. Ramakrishnan, Rev. Mod. Phys. $\mathbf{57}$,
287 (1985).


\newpage
\end{thebibliography}
\end{document}